\documentstyle[12pt, cite, epsfig, a4]{article}

\newcommand{\la}{\langle}
\newcommand{\ra}{\rangle}

\begin{document}

\hfill Nijmegen preprint

\hfill HEN-391

\hfill May 1996

\vspace{1cm}

\begin{center}

\bigskip
{\Large\bf Generalized Bunching Parameters and  Multiplicity \\
Fluctuations in Restricted Phase-Space Bins}                               %
\vspace{2.cm}

{\large S.V.Chekanov\footnote[1]{On leave from 
Institute of Physics,  AS of Belarus,
Skaryna av.70, Minsk 220072, Belarus},
W.Kittel}

\medskip

{\it High Energy Physics Institute Nijmegen
(HEFIN), University of Nijmegen/NIKHEF,\\
NL-6525 ED Nijmegen, The Netherlands}

\bigskip

{\large V.I.Kuvshinov} 

\medskip

{\it Institute of Physics,  AS of Belarus,
Skaryna av.70, Minsk 220072, Belarus}

\vspace{2.0cm}
\end{center}

\centerline{\bf ABSTRACT}

\vspace{0.5cm}
Experimental aspects of the use of bunching parameters
are discussed. 
Special attention is paid to the behavior  
expected for the case of purely statistical fluctuations.
We studied bin-averaged bunching parameters and
propose a generalization of bunching
parameters, making use of the interparticle distance-measure technique.
The proposed method opens up the possibility of carrying out 
a comprehensive and 
sensitive investigation of multiplicity fluctuations inside jets.

\vspace{2.0cm}
\centerline{(Submitted to Zeitscrift f\"{u}r Physik C)}

\newpage

\section { Introduction}

In recent years, multiparticle density fluctuations
have been studied in ever smaller  phase-space intervals $\delta$
in terms of normalized factorial
moments (NFMs) $F_q(\delta )$ \cite{zp1}. The NFMs can be defined as
\begin{equation}
F_q(\delta )=\frac{\sum^{\infty}_{n=q}n^{[q]}P_n(\delta )}
{\left(\sum^{\infty}_{n=1}nP_n(\delta )\right)^q}\, , 
\quad n^{[q]}=n(n-1)\ldots (n-q+1),
\label{1}
\end{equation}
where $n$ is the (charged) particle multiplicity and 
$P_n(\delta )$ is the multiplicity distribution in $\delta$.
The interval $\delta$ can be any interval in  phase space, such as 
in rapidity, azimuthal angle, transverse momentum, or a
(multi-dimensional) combination of these variables.
This method has recently been improved by the use of 
density and correlation integrals 
\cite{zp2} to avoid the problems of bin splitting and the 
insufficient use of
experimental statistics inherent to definition (\ref{1}). 

From an experimental point of view, the most important properties
of the NFMs are: 

1) they filter out Poissonian statistical noise;

2) events can contribute to (\ref{1}) only if $n\geq q$,
so they resolve the high-multiplicity tail of $P_n(\delta )$; 

3) if local self-similar dynamical multiplicity fluctuations
exist, then $F_q(\delta )\propto \delta^{-\phi_q}$, $\phi_q>0$. Such
a power-law behavior is called intermittency and the $\phi_q$ are called
intermittency indices. They are related to the anomalous
dimensions of the corresponding fractal 
system by the simple relation $d_q=\phi_q/(q-1)$.

Additional advantages of density integrals are that 
they avoid the problem
of bin splitting inherent to the definition of NFMs above, and
that they allow the use of general distance measures. 
Correlation integrals, furthermore, are based on genuine $q$-particle
correlations, which avoids trivial contributions from lower-order densities.
For reviews see \cite{z1,z2,zp3,zp4} and references therein.

Recently, another simple
mathematical tool has been proposed to investigate  multiparticle fluctuations.
In order to reveal 
intermittent structure of multiparticle production, 
it is, in fact, sufficient to study the 
behavior of the probability distribution near  multiplicity $n=q-1$
by means of the so-called bunching parameters (BPs) \cite{zp5,zp6}
\begin{equation}
\eta_q(\delta )=\frac{q}{q-1}\frac{P_q(\delta )P_{q-2}(\delta )}
{P_{q-1}^2(\delta )}.
\label{2}
\end{equation}
These quantities are formally identical to those used in quantum optics
\cite{opt}. The bunching-parameter 
method has also been extended to measure bin-bin correlations \cite{bibi}.

In the mathematical limit $\delta\to 0$,  the  
relation between NFMs and the BPs is
\begin{equation}
F_q(\delta )\simeq\prod_{i=2}^{q}\eta_i^{q-i+1}(\delta ).
\label{92}
\end{equation}
In this limit, therefore, the BPs share with the NFMs the important
property of suppression of  Poissonian
statistical noise. 

In fact, for an event sample following a Poissonian multiplicity
distribution, one finds $\eta_q(\delta )=1$ for all $q$ and $\delta $.
If all BPs are larger than $1$, the corresponding multiplicity
distribution is broader than the Poisson distribution.
On the other hand, a multiplicity distribution is narrower than  Poisson
if all its BPs are smaller than $1$.

For a sample of events with a fixed finite  number of particles $N$ in
full phase space, independent emission of these particles
leads to a (positive) binomial distribution in the interval
$\delta$. Consequently, the BPs have the 
values $\eta_q^{\mathrm{PBD}}=(q-1-N)/(q-2-N)$, i.e.,
are  again independent of $\delta$.

As shown in \cite{zp5}, there exists, in fact, a large class of 
multiplicity distributions for which the BPs are independent of $\delta$
for the full range of $\delta$ values. This result is the first important
point investigated in detail in this paper.

\medskip

The relevance of the bunching parameters for multiparticle production
in high-energy collisions, however, lies in the following properties:

\medskip

1) From (\ref{92}) we can see that the second-order BP follows  
$\eta_2(\delta )\sim\delta^{-\phi_2}$ 
for intermittent fluctuations
in the limit $\delta\to 0$ (bunching effect of the second order), 
while the  higher-order 
BPs may have any type of dependence on $\delta$ \cite{zp5}.
 
\medskip
   
2) In the case of  monofractal behavior, the
anomalous dimension $d_q$ is independent of $q$.
Variation of $d_q$ with increasing $q$
corresponds to a multifractal
behavior. In contrast to the NFMs, 
only $\eta_2(\delta )$ increases with decreasing $\delta$
for monofractal behavior, while the
$\eta_q(\delta )$ are constants for all $q>2$ \cite{zp5}. 
Any $\delta$ dependence of
higher-order BPs, therefore, reveals a deviation from monofractal 
behavior of the multiplicity fluctuation. 

\medskip
 
3) The lower-order BPs are more sensitive than the NFMs to spikes with
a small number of particles. 
Only spikes with $n\leq q$ particles can contribute to the 
bunching parameter of order $q$. Hence, the BPs act as  a filter,
but, in comparison to the NFMs, with a complementary property
(see property 2 of NFMs above). 

This feature of BPs is important for the study of high-multiplicity events,
where unusually large dips in the density distribution of individual events 
can be treated as a dynamical effect as well as that of 
the appearance of spikes. 
In this case, the lowest-order
BPs will be sensitive to such dips.
On the other hand, 
for lower-multiplicity reactions, 
such as $\mathrm{e}^+\mathrm{e}^-$-annihilation, 
the use of  BPs can provide high-precision
measurements of local fluctuations, 
since they  suffer less from the bias arising due to a
finite number of experimental events than do the NFMs 
(see property 6 below).

\medskip

4) The BPs have a more direct link than the NFMs to the 
multiplicity distribution itself \cite{zp5}. Any multiplicity
distribution can be expressed in terms of the BPs as  
\begin{equation} 
P_n(\delta )=P_0(\delta )\frac{\lambda^n(\delta )}{n!}
\prod^{n}_{i=2}\eta_i^{n-i+1}(\delta ),
 \qquad \lambda (\delta )=\frac{P_1(\delta)}
{P_0({\delta })}.
\label{111}
\end{equation} 
 
\medskip

5) From the theoretical point of view, the BPs are useful when direct
calculation of the NFMs from a model or theory becomes too tedious. 
Factorial moments are easily calculated from the generating
function of the multiplicity distribution. 
A large class of distributions exists, however, without  any 
simple analytical 
form of the generating function. 

\medskip
 
6) Moreover, from the experimental point of view, 
we expect that the BPs are less severely affected by 
the bias from finite statistics  
than are the NFMs:
In practice, the multiplicity distribution 
$P_n(\delta )$ is always truncated at large $n$ due to
finite statistics in a given experiment. 
As a consequence, the values of  high-order
NFMs at small bin size are determined  by the first few terms in
expression (\ref{1}) only, which leads in most cases  to a
significant underestimate of 
the  measured NFMs with respect to their true values \cite{zp7,zp77,zp8}.  
Furthermore, the calculation of a given-order BP
is simpler, since one is analyzing events for three given  
multiplicities only, without the requirement of normalization by 
an average multiplicity. 

\medskip

7) Another experimental advantage of the bunching-parameter 
measurements is that, for the calculation of the BP of order $q$,
one needs to know only the $q$-particle resolution of the detector.
In contrast, the precise calculation of the NFMs of order $q$ always 
involves the knowledge of the resolution of $n\geq q$ particles. 
So, for a given 
$q$-track resolution, the behavior of the $q$th-order NFM may contain
a systematic bias due to contributions from the tail of the
multiplicity distribution measured with insufficient resolution.

\medskip

The study of multiparticle production processes with the 
help of BPs, therefore, is expected to provide important information on
multiplicity fluctuations in ever smaller  phase-space intervals,
in addition to and complementary to that extracted with NFMs.

In Sect.~2, we discuss the problem of 
Poissonian noise and the behavior of
BPs for a number of theoretical models.
In Sect.~3, we give experimental definitions of the BPs
and suggest an extension of the
bunching-parameter method  to avoid the problem of bin splitting and
to allow a more general choice of distance measure, 
analogous to the extension of NFMs to the
density integrals mentioned above.
The crucial question of the
behavior of  BPs  and their extensions 
in the case of purely statistical phase-space fluctuations is shown in
Sect.~4. 
In Sect.~5, we give, as an example, 
a comparison of the factorial-moment and bunching-parameter analysis
of two different intermittent samples generated by the JETSET 7.4 model.

\section{Poissonian noise suppression and other properties}

\subsection{The problem of Poissonian noise}

As we noted in the introduction, the NFMs have an important feature
for the theoretical study of local fluctuations: 
they are not contaminated by
Poissonian statistical noise.
First, let us show that the BPs  reduce the statistical
noise in the limit $\delta\to 0$, as well,
meaning that
BPs are not only a convenient experimental tool that
can reduce the bias from finite statistics ($N_{\mathrm{ev}}\ne\infty$),
but also can suppress statistical noise arising due to
the finite number of particles per event ($N\ne\infty$). The last point is
of vital importance for the 
study of theoretical models with
an infinite number of particles in an event.

Let us first define a  particle density $\rho$ in bin $m$ for
an individual event as
\begin{equation}
\rho=\frac{n}{\delta},
\label{9lo}
\end{equation}
where $n$ is the number of particles in  bin $m$ of
size $\delta$.
For a local-fluctuation analysis,
we need to consider very small bin sizes, i.e., $\delta\to 0$.
$\rho$, therefore, is an asymptotic density, since it can be defined
in the limit of infinite multiplicity $N$ (or $n$) for a given event.

Using
another (theoretical) limit, $N_{\mathrm{ev}}\to\infty$,
we can define $\omega (\rho )$
as a  continuous probability density
to observe a given value of $\rho$.
This density
fulfills the normalization condition
\begin{equation}
\int_{0}^{\infty}\omega (\rho )\mathrm{d}\rho  =1.
\label{10lo}
\end{equation}

Of course, the density  $\rho$ for bin size $\delta$ fluctuates
around the average value
\begin{equation}
\la\rho\ra\equiv f_1=
\int_{0}^{\infty}\rho \>\omega(\rho )\mathrm{d}\rho .
\label{11lo}
\end{equation}
Because we are interested in the deviation of $\rho$ from the
average value $f_1$, the next step is to define the higher-order
moments of $w(\rho )$ as follows
\begin{equation}
\la\rho^q\ra\equiv f_q=\int_{0}^{\infty}\rho^q\omega(\rho )\mathrm{d}\rho .
\label{111lo}
\end{equation}

In experimental studies, the multiplicity $N$  is
finite. In this case, the number of particles $n$ in bin $m$
fluctuates around the average value
due to  ``statistical noise''.
If we accept this assumption, and the additional assumption that
such  a statistical noise does not introduce new fluctuations,
the observed (discrete) multiplicity distribution
$P_n(\delta)$ to observe $n$ particles in $\delta$ can be described by
the following Poisson transformation \cite{zp1}
\begin{equation}
P_n(\delta )=\int_{0}^{\infty}\omega (\rho )
\frac{(\rho\delta )^n \exp (-\rho\delta )}{n!}\mathrm{d}\rho.
\label{12lo}
\end{equation}
Expression (\ref{12lo}) represents a convolution of the
statistical Poissonian noise of mean  $\rho\delta$
with a true, dynamical distribution $\omega (\rho )$.

The next problem, therefore, is how to compare model
fluctuations described by $\omega(\rho )$
with the experimental fluctuations
defined by $P_n(\delta)$.
Substituting (\ref{12lo}) in the definition
of factorial moments gives
\begin{equation}
\la n^{[q]}\ra\equiv\sum_{n=q}^{\infty}n^{[q]}P_n(\delta ) = \delta^q f_q,
\qquad q=1,2,3\ldots ,
\label{14lo}
\end{equation}
where $f_q$ are ordinary moments defined by (\ref{11lo}) and (\ref{111lo}).
Hence, for NFMs (\ref{1}) one obtains
\begin{equation}
F_q(\delta )=\frac{f_q}{f_1^q}.
\label{16lo}
\end{equation}
The right side of this expression represents the normalized moments given
by a model  distribution $\omega (\rho )$.
Studying this distribution in experiments with finite
$N$, therefore, is equivalent to measuring the  NFMs $F_q(\delta )$.

Let us note that in the limit of small phase-space size, we can 
only keep the leading term in expression
(\ref{12lo}), i.e., $P_n(\delta )$ can be rewritten as
\begin{equation}
P_n(\delta )\simeq\frac{\delta^n}{n!}
\int_{0}^{\infty}\omega (\rho )\rho^n\mathrm{d}\rho
\label{23lo}
\end{equation}
if fluctuations in a model are investigated in the limit $\delta\to 0$.
Substituting this expression in ({\ref{2}}) gives
\begin{equation}
\eta_q(\delta )\simeq\frac{f_qf_{q-2}}{f_{q-1}^2},
\label{24lo}
\end{equation}
where $f_0=1$ according to (\ref{10lo}) and (\ref{111lo}).
Therefore, $\eta_q(\delta )$ calculated from  experiment gives 
information on the fluctuations
described by the theoretical probability density $w(\rho )$, since
Poissonian contributions cancel at small $\delta$.
From (\ref{24lo}) and (\ref{16lo}) one can obtain relation (\ref{92})
given in the introduction.

The idea to express intermittency directly in terms of the 
probabilities has also been proposed by Van Hove \cite{hove}.
Indeed, in the limit $\delta\to 0$, one can use the ratio
$P_q(\delta )/P^q_1(\delta )$ instead of $F_q(\delta)$, since
\begin{equation}
\frac{P_q(\delta )}{P_1^q(\delta )}\propto\frac{f_q}{f_1^q}=F_q(\delta),
\label{244lo}
\end{equation}
according to (\ref{23lo}).

\subsection{Multifractal and monofractal behavior}

For a model with intermittent behavior, we expect
\begin{equation}
\frac{f_q}{f_1^q}\propto\delta^{-\phi_q}.
\label{25lo}
\end{equation}
Using this relation and (\ref{24lo}), one obtains 
\begin{equation}
\eta_q(\delta )\propto\delta^{ 2\phi_{q-1}-\phi_q -\phi_{q-2}},
\qquad \delta\to 0 ,
\label{26lo}
\end{equation}
where $\phi_0 =\phi_1 =0$.

As a reminder, one should expect
$\phi_q=d_2(q-1)$ for monofractality.
For these types of fluctuations, the BPs  have the following behavior
\begin{equation}
\eta_2(\delta )\propto\delta^{-d_2}, \qquad
\eta_{q>2}(\delta )\simeq \mbox{const}.
\label{27lo}
\end{equation}
For monofractal behavior, therefore,
what one obtains is that all high-order BPs $\eta_{q>2}(\delta )$
are $\delta$-independent constants.
This result is one of the  important advantages
of the bunching-parameter method over factorial
moments: to reveal multifractal
behavior in an experimental sample, it is not necessary
to interpolate an experimental slope by the power-law
$F_q(\delta )\propto\delta^{-d_q(q-1)}$
in order to derive a $q$-dependence of $d_q$.

\subsection{Examples}

For illustrative purposes, we now consider  examples
of the behavior of BPs
for various dynamical models:

\subsubsection{Random-cascade model}

This is the first model \cite{zp1} used in
high-energy fluctuation phenomenology.
For this model, the intermittency indices
have the following form
\begin{equation}
\phi_q=\frac{1}{2}q(q-1)d_2.
\label{28lo}
\end{equation}
From (\ref{26lo}), one can see that all BPs follow the same power law
\begin{equation}
\eta_q(\delta )\propto\delta^{-d_2}, \qquad \mbox{for all}
\quad  q\ge 2.
\label{29lo}
\end{equation}
This feature in the behavior of the cascade model can be revealed
by calculating the BPs and by comparing their 
power-law behavior, without
the necessity of any fit of NFMs by a power-law.

\subsubsection{Second-order phase transition}

One expects \cite{zp9} that for a system 
undergoing a second-order phase transition
the corresponding intermittency indices would depend linearly
on the rank of the moment
\begin{equation}
\phi_q=d_2(q-1).
\label{30lo}
\end{equation}
Such a behavior has been derived from a 
toy Ising model \cite{zp99a,zp99b,zp99c}.
In this case, according to (\ref{27lo}), all higher-order BPs are 
$\delta$-independent constants.

\subsubsection{Perturbative QCD cascade}
 
In a QCD cascade with fixed coupling
constant $\alpha_s$,  the  intermittency 
indices have the following multifractal 
behavior \cite{zp10,zp11} 
\begin{equation}
\phi_q=D(q-1)-\gamma_0 r_q, \qquad r_q=(q-1)(q+1)q^{-1},
\label{31lo}
\end{equation}
where $D$ is the topological 
dimension of the phase space 
under consideration and $\gamma_0=(6\alpha_s/\pi )^{1/2}$
is the QCD anomalous dimension.
From (\ref{26lo}), one can conclude
that the behavior of all  high-order
BPs is $D$-independent
for a fixed-coupling regime of
QCD and is governed only by $\gamma_0$
\begin{equation}
\eta_q(\delta )\propto\delta^{\gamma_0 h_q},
\qquad h_q=r_q + r_{q-2} - 2r_{q-1}, \qquad q\ge 3,
\label{32lo}
\end{equation}
where $r_1=0$.
As a first rough test of the QCD prediction, therefore,
a measurement of the third-order BP for
different dimensions $D$ can provide a qualitative
answer to the applicability of this type of QCD calculations to
real data. Note that this can be done very precisely, since
statistical (and systematical) errors are small
for a third-order BP.

\section{Experimental definitions of BPs}

\subsection {Bin-averaged BPs}

In order to  increase the statistics and to reduce 
the statistical error of observed BPs when analyzing experimental data,
we can use  bin-averaged BPs as defined in
analogy to the bin-averaged factorial moments:

\medskip

1) {\em Flat phase-space distribution:} 
The following definition of horizontally normalized 
bin-averaged BPs can be used \cite{zp5}: 

\begin{equation} 
\eta^{\mathrm{hor}}_q(\delta )=
\frac{q}{q-1} \frac{\bar N_q(\delta ) \bar N_{q-2}(\delta )}
{\bar N_{q-1}^2(\delta )} \quad,
\label{3}
\end{equation}
where 
\begin{equation} 
\bar N_{q}(\delta )=\frac{1}{M}\sum_{m=1}^{M}N_q(m,\delta ) \quad.
\label{4}
\end{equation}
Here, $N_q(m,\delta )$ is the number of events  having $q$ particles
in  bin $m$, $M=\Delta /\delta$  is the total number of bins, and
$\Delta$ represents the full phase-space volume. 

\medskip

2) {\em Non-flat phase-space distribution:}
In this case we need to use
vertically normalized  BPs defined as 
\begin{equation} 
\eta^{\mathrm{ver}}_q(\delta )=M^{-1}\frac{q}{q-1}
\sum_{m=1}^{M}\frac{N_q(m,\delta )
 N_{q-2}(m,\delta )}{N_{q-1}^2(m,\delta )} \quad.
\label{5}
\end{equation}

It should be pointed out that, in this case, the sum  runs  over
non-zero bins only. This type of BPs, therefore,
demands more statistics and
may be unstable for small phase-space bins. In contrast,
events with  no particles in a bin can contribute to 
the horizontally normalized  BPs.
For this reason, it may be 
more convenient to use the BPs (\ref{3}) for non-flat distributions as well.
To be able to do this, one must carry out a transformation from the original   
phase-space  variable to one
in which the underlying distribution is approximately
uniform \cite{zp12,zp13}.

\subsection{Generalized distance measure}
\subsubsection{Definitions of spike size}

The main deficiency of  definitions (\ref{3}) and (\ref{5})
(and the bin-averaged NFMs)  lies
in the artificial splitting of particle spikes.
Spikes do not contribute to the  $N_q(m,\delta )$ 
if the boundaries of bins happen to split such spikes.
This deficiency can be avoided 
by the choice of a proper
distance $X_{i,j}$ between two particles, which as
demonstrated in \cite{zp14}, would have  the additional 
advantage of largely increasing the statistics 
effectively used in a given experiment, at a given resolution.

For a given event, let us define a $g$-particle spike of size $\epsilon$
as a group of $g$ particles having mutual phase-space distance $X_{i,j}$
smaller than  $\epsilon$. According to this definition,
the condition for  particles to belong  to  a spike is
\begin{equation} 
\prod^g_{i=1}\prod^g_{j=1}\theta (\epsilon -X_{i,j})=1,
\qquad i\ne j,
\label{55}
\end{equation}
where $\theta$ is the Heaviside unit step function.
To determine the spike size $\epsilon$ for a given event 
we  have  used here
the so-called Grassberger-Hentschel-Procaccia (GHP)
counting topology
\cite{zp15,zp16}, for which a $g$-particle hyper-tube is assigned
a size $\epsilon$ that corresponds to the
maximum of all pairwise distances.

Alternative topologies are 
the so-called ``snake'' topology \cite{zp2} 
\begin{equation} 
\prod^g_{i=2}\theta (\epsilon -X_{i-1,i})=1,
\label{515}
\end{equation}
which corresponds to the longest distance between two
particles connected by one joining line, and 
the ``star''  topology \cite{zp14} defined as
\begin{equation} 
\prod^g_{i=2}\theta (\epsilon -X_{1,i})=1.
\label{525}
\end{equation}
The star topology involves all 
particles that are paired with a preselected center
particle (index $1$). 
It shares all the advantages of the GHP and snake forms, and
is computationally more efficient.

\subsubsection{Bunching parameters}

After establishing the definitions of spike size $\epsilon$, 
we can investigate the behavior of
multiplicity fluctuations in ever smaller $\epsilon$ 
by means of the bunching-parameter method.

\medskip
{\em Differential BPs:}
\medskip

In any multiparticle process, the
number of $g$-particle spikes fluctuates around 
an average value according to a certain 
probability distribution. Let $P_n(\epsilon ,g)$ be
the probability distribution of observing in an event 
a number $n$ of
$g$-particle spikes of size $\epsilon$, 
irrespective of the presence of other spikes.
This distribution can be  characterized by the generating function 
$G(\epsilon ,g)$ 
defined as
\begin{equation} 
G(z, \epsilon ,g)=\sum^{\infty}_{n=0}P_n(\epsilon ,g)z^n.
\label{1we}
\end{equation}
For a purely independent production of spikes, 
the multiplicity distribution $P_n(\epsilon ,g)$  
follows a Poissonian law, 
\begin{equation} 
P_n^{\mathrm{P}}(\epsilon ,g)=(n!)^{-1}\bar{K}^n(\epsilon ,g)
\mathrm{e}^{-\bar{K}(\epsilon ,g)},
\label{1wwe}
\end{equation}
with a  generating function of the form
\begin{equation} 
G^{\mathrm{P}}(z, \epsilon ,g)=
\mathrm{e}^{\bar{K}(\epsilon ,g)(z-1)},
\label{2we}
\end{equation}
where $\bar{K}(\epsilon ,g)$ represents the average number of $g$-particle
spikes of size $\epsilon$ in an event in the sample under study.

To measure the distribution $P_n(\epsilon ,g)$  without
the contribution from  
events with a large number of such spikes 
(or ``tail'' of the real distribution), one can calculate
the following ``differential'' type of BPs
\begin{equation} 
\chi_{q}(\epsilon ,g )=
\frac{q}{q-1} \frac{\Pi_q(\epsilon , g) \Pi_{q-2}(\epsilon ,g)}
{\Pi^2_{q-1}(\epsilon , g)}, \quad q=2,3,\ldots , 
\label{3we}
\end{equation}
where $\Pi_i(\epsilon ,g)$ represents the number of events with
a number $i$ of $g$-particle  spikes of size $\epsilon$. 
For purely independent emission of spikes,
$P_n(\epsilon ,g)$ follows the Poissonian
distribution (\ref{1wwe}) and  all BPs (\ref{3we}) 
are equal to unity for all $q$ and $\epsilon$.

\medskip
{\em Integral BPs:}
\medskip

Of course, when 
analyzing experimental data, it is  difficult to obtain all values of
$\chi_{q}(\epsilon ,g )$ as a function of $\epsilon$. This
is due to the large number ($=q\,g$) of possible configurations involved and
the finite number of events available.
We can, however, use a less 
informative and less differential definition suitable for an
experiment with rather small statistics. 

To understand these kinds of measurements, let us
first define the probability distribution $P_n(\epsilon )$ to
observe in an event 
a number $n$ of multiparticle spikes, irrespective of
how many particles are inside each spike. 
From a theoretical point of view, if  all $g$-particle spikes
are produced independently of each other, the generating function
$G(z, \epsilon)$ for  $P_n(\epsilon )$ has the form of a convolution
of spike distributions with different particle content, i.e., 
\begin{equation} 
G(z, \epsilon)=\prod^{\infty}_{g=2}G(z, \epsilon, g).
\label{4we}
\end{equation}

For purely independent spike production, one has
from (\ref{2we}) and (\ref{4we}), 
again a Poissonian distribution, with the generating function 
\begin{equation} 
G(z, \epsilon)=G^{\mathrm{P}}(z, \epsilon)=
\mathrm{e}^{\bar{K}(\epsilon )(z-1)}
\label{5we}
\end{equation}
and with the average number of multiparticle spikes
\begin{equation} 
\bar{K}(\epsilon )=\sum_{g=2}^{\infty}\bar{K}(\epsilon ,g).
\label{6we}
\end{equation}
As mentioned before, 
to measure a deviation from the Poissonian distribution,
one can calculate the ``integral'' type of BPs
\begin{equation} 
\chi_{q}(\epsilon )=
\frac{q}{q-1} \frac{\Pi_q(\epsilon ) \Pi_{q-2}(\epsilon )}
{\Pi^2_{q-1}(\epsilon )}, \quad q=2,3,\ldots , 
\label{7we}
\end{equation}
where $\Pi_i (\epsilon )$ represents the number 
of events with  $i$  spikes of size $\epsilon$, irrespective of
how many particles are inside each spike. 
If $\chi_{q}(\epsilon )\ne 1$, then the conclusion of 
non-Poissonian spike production follows and a more sophisticated analysis
can be performed with the help of the differential kind of BPs.

According to the definition above, all spikes with $g\ge 2$ 
particles contribute to $\chi_{q}(\epsilon )$. However, 
one can propose
a more selective study of the spike fluctuations. Indeed, in the
case of purely random (Poisson) fluctuations, the probability 
distributions
to observe $n$ spikes with $g\ge s$ or with $g\le s$ particles
($s$ is some integer number) also follow the Poissonian law
due to the ``reproductive'' property of the Poisson distribution.
In terms of generating functions, these two distributions
can be expressed as
\begin{equation} 
G(z, \epsilon, g\ge s)=\prod^{\infty}_{g=s}G(z, \epsilon, g)=
\exp \left[\sum_{g=s}^{\infty}\bar{K}(\epsilon ,g)(z-1)\right]
\label{8we}
\end{equation}
and
\begin{equation} 
G(z, \epsilon, g\le s)=\prod^{s}_{g=2}G(z, \epsilon, g)=
\exp \left[\sum_{g=2}^{s}\bar{K}(\epsilon ,g)(z-1)\right].
\label{9we}
\end{equation}
To measure a deviation from these distributions,  
instead of $\Pi_i (\epsilon )$,
one must use in (\ref{7we}) the number of events $\Pi_i(\epsilon ,g\ge s)$
and $\Pi_i(\epsilon ,g\le s)$  having  $i$ spikes with
$g\ge s$ and $g\le s$ particles, respectively.  
The definition with $\Pi_i(\epsilon ,g\le s)$
is more preferable for high-precision measurements,
because this quantity does not contain the contributions 
from spikes with high-multiplicity content.

\medskip
{\em Discussion:}
\medskip

The main reason for introducing 
the integral BPs (\ref{7we}) 
is that the $\chi_{q}(\epsilon )$ are more useful
when the statistics of an experiment are small. 
In this case, the lower-order BPs (\ref{3we}) 
have large statistical errors
\footnote{According to the Gauss law, the statistical error
on the  number of events $\Pi$ 
is $\sqrt{\Pi}$ for large  $\Pi$.},
whereas higher-order BPs even vanish.
In contrast, the  BPs  (\ref{7we})
have smaller statistical errors and high-order BPs
can be still calculable.
Moreover, the simplicity of this definition  
makes the latter very economical to
calculate.

The actual choice of the definition of the BPs 
and of the value of $\epsilon$ strongly depends  on
the aims of the specific investigation. 
For example, at {\em large} $\epsilon$ 
the BPs  are sensitive to the large scale of
an event structure, where any jet behaves as a cluster (a spike
of dynamical origin). The calculation of the BPs
according to (\ref{7we}), therefore, 
corresponds to a study of a fluctuation in the 
number of jets, where each jet is considered, 
regardless of its inner structure. 
For an intermittent fluctuation, 
we expect that all second-order BPs are a power-like function of 
$\epsilon$ for $\epsilon\to 0$, whereas high-order ones 
can have any dependence on $\epsilon$.

All these kinds of definitions  
have an important advantage over the conventional definition
(\ref{3}) or (\ref{5}):   
we now can study the structure of spike fluctuations.
In addition, we can investigate a given sample
in a variety of new variables. 
For example, the squared four-momentum difference
between any two particles $Q^2_{12}=-(p_1-p_2)^2$ is 
theoretically preferred
for investigations of Bose-Einstein or effective mass correlations. 

The question remains  why 
we use the definitions of the generalized
BPs in terms of the spike multiplicity distributions $P_n(\epsilon ,g)$
and $P_n(\epsilon )$. Indeed, at first sight, it may seem  more
straightforward to use a conventional probability
$\tilde{P_n}(\epsilon )$ of having  $n$ particles
inside a hyper-tube of size $\epsilon$. 
This probability can be found as
\begin{equation}
\tilde{P_n}(\epsilon)=\frac{K_n(\epsilon)}{\sum_{n=1}^{\infty}K_n},
\label{1zpo}
\end{equation}
where $K_n(\epsilon)$ is the number of $n$-particle spikes 
(hyper-tubes) of size $\epsilon$ found in $N_{\mathrm{ev}}\to\infty$ 
experimental
events. Clearly,  $\tilde{P_0}(\epsilon )$ does not exist.
Hence, the BPs
\begin{equation}
\eta_q(\epsilon )=\frac{q}{q-1}\frac{K_q(\epsilon )K_{q-2}(\epsilon )}
{K_{q-1}^2(\epsilon )}
\label{2zpo}
\end{equation}
exist only for $q=3,4,\ldots$, but not for $q=2$. 
It is important to note, however, 
that $\tilde{P_n}(\epsilon )$ is not Poissonian 
even if particles are distributed independently
(see Fig.~4 and the comments in Sect.~4.2)
\footnote{ 
Such a non-Poissonian form of  $\tilde{P_n}(\epsilon )$
has also been realized in \cite{zp14}, where a complex event-mixing
technique has been introduced to normalize generalized factorial moments.}. 
In addition,
we will show that $\eta_q(\epsilon )$ suffers from  insufficient
statistics. Of course, 
if we keep both these problems in mind, the $\eta_q(\epsilon )$
can be used for experimental study as well.

Note that for the generalized BPs (\ref{3we}) and  (\ref{7we})
we use the letter $\chi_q$ in order to emphasize that these
definitions are intended for measuring of the bunching of spikes,
rather than that of particles. From this point of view, no simple connection
exists between $\eta_q(\delta )$ (or $\eta_q(\epsilon )$) and
$\chi_q(\epsilon)$. The same is true for the conventional and
the generalized NFMs \cite{zp14}. 
Furthermore, the relation between the NFMs  and
the BPs $\chi_q(\epsilon)$ 
ceases to have a 
simple form. As the result, it is no longer possible
to draw a  conclusion on the 
$\epsilon$-dependence of  the $\chi_q(\epsilon)$
from the study of the generalized NFMs. 
The question of the relation between the generalized BPs and the generalized 
NFMs
will be  the subject of a future paper. Below, 
we will, however,  demonstrate that, as is the case for the NFMs, 
a rise of the value of 
$\chi_q(\epsilon )$ with decreasing $\epsilon$ is inherent in realistic
systems exhibiting intermittency.

Unfortunately, the problem of purely random (or statistical) fluctuations
cannot always be reduced  to the study of Poissonian distributions.
Below, we will consider
a general case of phase-space statistical fluctuations for
which the property $\chi_{q}(\epsilon ,g )=1$, $\chi_{q}(\epsilon )=1$
is only a particular case, corresponding to a full-phase-space
Poissonian multiplicity distribution.

\subsubsection{Propagation of the statistical error for generalized BPs}

As is the case for the extension of the usual NFMs to the
density integrals, the estimation of the statistical error is simplified for
generalized,  as compared to,  bin-averaged BPs. The calculation of
the statistical error (i.e. the standard deviation) for the BPs
(\ref{3}) and (\ref{5}) includes bin-bin correlation coefficients  
(all $M$ bins are dynamically correlated)  
not present in the other definitions.

In the following, we derive
an exact expression for the standard deviation of the generalized BPs
using a distance measure $\epsilon$.
For simplicity, we shall use the symbolic expression 
\begin{equation}
\chi_{q}=\frac{q}{q-1} \frac{ \Pi_{q}
\Pi_{q-2}}
{\Pi_{q-1}^2},
\label{105}
\end{equation}
where $\Pi_q$  stands for any definition of
the number of events having a given 
{\em spike configuration $q$} used in (\ref{3we}) and (\ref{7we}).

Let $W_q(t)$ be an indicator for the
presence of a given spike configuration (index $q$) in an experimental event
(integer argument $t$), i.e., for a given measurement $t$ we set
\begin{equation}
W_q(t)=
\left\{ \begin{array}{ll} 1, & \mbox {if spike configuration $q$ is occuring,} \\
0, & \mbox {otherwise.}
\end{array}
\right.
\label{116}
\end{equation} 
After $N_{\mathrm{ev}}$ measurements, we get  the sample mean of $W_q(t)$
\begin{equation}
{\overline W}_q=
\frac{\sum^{N_{\mathrm{ev}}}_{t=1}W_q(t)}{N_{\mathrm{ev}}}=
\frac{\Pi_q}{N_{\mathrm{ev}}}\, .
\label{154}
\end{equation} 

It can be seen that the definition of generalized BPs
(\ref{105}) already represents an average value 
\footnote{Here we applied the fundamental statistical
assumption that, to a first approximation, $\bar V=V(\bar x)$, where
$V(x)$ is a function of the directly measured quantity $x$.}
of BPs after $N_{\mathrm{ev}}$
measurements with the sample mean ${\overline W}_q$, since
$N_{\mathrm{ev}}^2$  cancels in definition (\ref{105}).
Let us note that all our BPs exist only as an average quantity,
since we do not use any
definition for BPs with  $W_q(t)$ for a single experimental 
event. 

The elements of the covariance matrix for an unbiased estimator are
given by the standard expression 
\begin{equation}
V_{q,q'}=\frac{1}{N_{\mathrm{ev}}(N_{\mathrm{ev}}-1)}
\left[\sum^{N_{\mathrm{ev}}}_{t=1}W_q(t)W_{q'}(t) - 
N_{\mathrm{ev}}{\overline W}_q{\overline W}_{q'}\right].
\label{107}
\end{equation} 
For $q=q'$, the covariance matrix reduces to the unbiased sample
variance $s_q^2$
\begin{equation}
V_{q,q}=s_q^2.
\label{108}
\end{equation} 

Given the covariance matrix, we can obtain
the sample variance $S_q^2$ for the generalized BPs using a general 
rule for combining correlated errors \cite{zp17},
\begin{equation}
S_q^2=\left(\frac{q}{q-1}\right)^2\left[\frac{{\overline W}_{q-2}^2}
{{\overline W}_{q-1}^4}
s^2_q + \frac{4{\overline W}_{q}^2{\overline W}_{q-2}^2}{{\overline W}_{q-1}^6}s_{q-1}^2 +
\frac{{\overline W}_{q}^2}{{\overline W}_{q-1}^4}s_{q-2}^2 +\varrho_q\right],
\label{159}
\end{equation} 
where $\varrho_q$ is a function of non-diagonal elements of the
covariance matrix describing  the correlations between the $W_q$,
\begin{equation}
\varrho_q=2\,V_{q,q-2}\frac{{\overline W}_{q}{\overline W}_{q-2}}{{\overline W}_{q-1}^4}
 - 4\,V_{q,q-1}\frac{{\overline W}_{q}{\overline W}_{q-2}^2}{{\overline W}_{q-1}^5} -
4\,V_{q-1,q-2}\frac{{\overline W}_{q}^2{\overline W}_{q-2}}{{\overline W}_{q-1}^5}.
\label{160}
\end{equation} 
The standard deviation is the square root of
the variance (\ref{159}). Let us note that for the calculation of
the standard deviation we did not use any assumption on 
a Gaussian distribution of  $W_q$. In fact, a Gaussian distribution
is, in general, not  applicable for the calculation of
statistical errors for  small $\epsilon$. The errors plotted in
the forthcoming figures are the errors calculated according to (\ref{159}).

\section{Statistical fluctuations and BPs}

As was shown in Sect.~2, BPs are not affected by
Poissonian noise in the limit $\delta\to 0$. However,
in order to use the BPs to extract information on
dynamical fluctuations, one has to know
their behavior in the case of purely
random phase-space fluctuations for realistic 
values of $\delta$.

The random fluctuations cannot always 
be described in terms of a  Poissonian distribution,
since  in multiparticle experiments, 
the full-phase-space multiplicity distribution
is often far from Poissonian. In addition, there is always a constraint
on the maximum  value of multiplicity 
because of  energy conservation.
This constraint can lead to non-Poissonian fluctuations in 
small phase-space intervals, even if the particles are produced in
phase space randomly, without any dynamical correlations. 

To study  statistical fluctuations, 
therefore, we consider a {\em general} case of
independent particle emission, when  
spikes appearing in phase space 
are caused by random properties
of an experimental sample.

\subsection{The bin-averaged BPs}

\subsubsection{Flat phase-space distribution}

In order to understand the  behavior of BPs (\ref{3}) and (\ref{5}) 
in the case of purely statistical fluctuations, we start with a
phase-space  distribution which is flat and equally wide 
for all multiplicities $N$.
In this case, the  
number $N_q(m,\delta )$ of events having $q$ particles in bin $m$ 
does not depend on the position of the bin, i.e., 
$N_q(m,\delta )=N_q(\delta )$.
Expressions  (\ref{3}) and (\ref{5}), therefore, are reduced to (\ref{2}).

An event sample with {\em purely statistical fluctuations in 
restricted phase space}
can be described by the following expression \cite{zp19,zp20,zp21} :
\begin{equation}
P_n^{\mathrm{stat}}(\delta )=\sum_{N=n}^{\infty}P_N
C^n_Np^n(1-p)^{N-n}, \quad p=\frac{\delta }{\Delta },  
\label{9}
\end{equation}
where $P_N$ is the multiplicity distribution for full phase space,  
the $C^n_N$ are the binomial coefficients 
and $p$ is the probability that a 
particle falls within a given interval $\delta$.
Expression (\ref{9}) states that for each data subsample
of events with fixed finite multiplicity $N$, particles
fall into  $\delta$ independently, i.e., according to
a (positive) binomial distribution \cite{zp22}.

When we speak of  purely statistical phase-space fluctuations in the case of
a finite number of particles in a single event, we imply 
independent emission of the particles into a small phase-space 
interval, i.e., without any interaction between particles
yielding dynamical spikes or clusters.
Of course, for a single event, even independent
emission can produce spikes, but only of  statistical nature.
In such a case as this, a multiplicity distribution obtained
after $N_{\mathrm{ev}}\to\infty$ 
experimental measurements can be expressed in 
the form of (\ref{9}).

Let us note that the statistical fluctuations described by (\ref{9})
have nothing to do with statistical noise described by 
Poisson transformation (\ref{12lo}). The notion of  statistical noise 
is necessary to take into account 
the finiteness of the number of particles in 
the counting bin (and, hence, in  full phase space).
We can get an ``observed'' discrete multiplicity
distribution from a  ``true'' continuous dynamical probability
density using the so-called linear transformation (\ref{12lo}) of the
density with a Poisson kernel.

Let
\begin{equation}
G(z,\delta )=\sum_{n=0}^{\infty }P_n(\delta )z^n
\label{91}
\end{equation}
be the generating
function  for the multiplicity distribution  $P_n(\delta )$ for $n$
particles in a small phase-space interval $\delta\leq\Delta $.
Then, if we multiply (\ref{9}) by $z^n$ and sum the 
result over $n$, we can find the
generating function for $P_n(\delta )$ as follows:
\begin{equation} 
G^{\mathrm{stat}}(z,\delta )=
\sum^{\infty}_{N=0}P_N(pz-p+1)^N.
\label{8}
\end{equation}
Using the relation between factorial moments
and  generating function 
\begin{equation} 
\langle n^{[q]}\rangle 
=G^{(q)}(z)\mid_{z=1}\, ,
\label{83}
\end{equation}
one finds
that the NFMs for  distribution (\ref{8}) 
are $\delta$-independent constants \cite{zp23} of 
the  form \cite{zp20}
\begin{equation}
F^{\mathrm{stat}}_q(\delta )=\frac{\left<N^{[q]}\right>_N}{\left<N\right>^q_N},   
\label{10}
\end{equation}
where $\langle\ldots \rangle$ denotes the average over all events 
following the probability distribution $P_N$:
\begin{equation}
\left<N^{[q]}\right>_N=\sum^{\infty}_{N=0}P_NN^{[q]}, \qquad q=1,2\ldots   
\label{7}
\end{equation}
Using  definition (\ref{2}), of the BPs, together with (\ref{9}), 
we obtain the BPs for the case of  purely statistical fluctuations 
\begin{equation}
\eta^{\mathrm{stat}}_q(\delta )=\frac{B_q}{B_{q-1}}, \qquad
B_q=\frac{\sum_{N=0}^{\infty}P_N N^{[q]}(1-p)^N}
{\sum_{N=0}^{\infty}P_N N^{[q-1]}(1-p)^N}.   
\label{11}
\end{equation}
If the phase-space interval is small enough,
then $(1-p)\to 1$ and (\ref{11}) is reduced to
\begin{equation}
\eta^{\mathrm{stat}}_q(\delta )\to
\frac{\left<N^{[q]}\right>_N \left<N^{[q-2]}\right>_N}    
{\left<N^{[q-1]}\right>_N^2},
\qquad  \delta\to 0,  
\label{12}
\end{equation} 
i.e., the BPs become independent of $\delta$.

If the multiplicity $N$ 
for full phase space follows a Poissonian distribution
with the average multiplicity $\bar{N}$,
then the corresponding generating function has the form
\begin{equation}
G^{\mathrm{P}}(z)=\mathrm{e}^{\bar{N}(z-1)},
\label{12ss}
\end{equation}
and (\ref{8}) again leads to the generating function for a
Poissonian distribution in
a small bin $\delta$
\begin{equation}
G^{\mathrm{stat}}(z)=\mathrm{e}^{p \, \bar{N}(z-1)},
\qquad  p=\frac{\delta}{\Delta} .
\label{13ss}
\end{equation}
In this case, the values of all-order BPs are unity for all $\delta$.
However, in many experiments $P_N$ is far from the Poissonian distribution,
and an additional study of the behavior of BPs for
purely statistical phase-space fluctuations is necessary. 

As an example, we present in Fig.~1 the behavior of the BPs as a function of
$M=\Delta /\delta$ for the case of 
statistical fluctuations according to
(\ref{9}) with
a truncated full-phase-space multiplicity distribution $P_N$ obtained
from the Monte-Carlo event generator JETSET 7.4 PS  \cite{zp23a} simulating the 
decay of a $\mathrm{Z^0}$. 
The generator was tuned according to the parameter
set of L3 Collaboration \cite{zp23b}.
The number of events generated is  
$750$k. In this sample, $P_N=0$ for $N<4$ and $N>70$ 
due to limited statistics.  
Let us stress that  we are using the analytical expression (\ref{11}), 
together with the $P_N$ simulated 
for full phase space from JETSET 7.4 PS, where $P_N$ is not equal, but   
similar, to a negative-binomial distribution with the
average charged-particle multiplicity $\bar{N}\simeq 21$.

As can be seen we from Fig.~1, the values of the BPs are larger than unity, but 
the approximation
$\eta_q^{\mathrm{stat}}(\delta )\simeq const$ for $M>10 -20$ will be a  good
estimate of the statistical fluctuations in 
an experimental situation where $P_N$ for 
full phase space is close to a truncated
negative-binomial distribution.  
For intermittent fluctuations, as a rule, we need to study
the behavior of the NFMs for much larger $M$. For such a situation, 
any observed dependence of the BPs (\ref{3}) on the interval size must be
caused by dynamical fluctuations.


\subsubsection{Non-flat phase-space distribution}


In the case of a non-flat phase-space distribution, the parameter $p$
becomes a function of  $N$, $\delta$, and the position of the bin
in phase space. Mathematically, this can be written as \cite{zp19}
\begin{equation}
p_m(N,\delta )=\frac{\int^{\delta_{m}}_{\delta_{m-1}}
\frac{\mathrm{d}N}{\mathrm{d}\delta}\mathrm{d}\delta }{\int_{\Delta}
\frac{\mathrm{d}N}{\mathrm{d}\delta}\mathrm{d}\delta } \quad ,   
\label{81}
\end{equation}
where the  phase-space density $\mathrm{d}N/\mathrm{d}\delta$ is defined for
a large set of events with a fixed total multiplicity $N$.
For small $\delta$ and non-singular 
phase-space density, each term in the sum (\ref{5})
is $\delta$-independent according to (\ref{12}) and, again, one has 
$\eta_q^{\mathrm{ver}}(\delta )\simeq const.$ 


\subsubsection{Theoretical aspect of the problem} 


From the theoretical point of view,
there is a class of distributions, $P_N$, for which the BPs are
$\delta$-independent constants, also for large $\delta$.
Let $G^{\mathrm{full}}(z)$ be 
the generating function for $P_N$ in full phase space.
After the composition with the positive-binomial distribution 
according to (\ref{8}),
the $G^{\mathrm{full}}(z)$ becomes 
$G^{\mathrm{stat}}(z,\delta )=G^{\mathrm{full}}(pz-p+1,\delta )$.
Then, the BPs
will be $\delta$-independent if the generating 
function $G^{\mathrm{full}}(pz-p+1, \delta )$ can be  expressed as 
\begin{equation}
G^{\mathrm{full}}(pz-p+1, \delta )=G^{\mathrm{full}}(1-p, \delta ) 
Q(z\lambda (\delta ) ),
\label{13}
\end{equation} 
where $Q(z\lambda (\delta ))$ is some function containing only
the combinations $z\lambda (\delta )$
(see (\ref{111}), where $\lambda (\delta )$ is a function of $\delta $).
Here, $G^{\mathrm{full}}(1-p,\delta )$ is equal to  
$G^{\mathrm{full}}(pz-p+1, \delta )$ for $z=0$. Expression (\ref{13})
can be  obtained from (\ref{111}) 
by setting $\eta_q(\delta )=const$ \cite{zp5}. 

If the multiplicity distribution
for full phase space is Poisson, binomial,
geometric, logarithmic, or negative binomial, then  the BPs do not
depend on $\delta $, even if  $\delta$ is not small \cite{zp5}.

As an example, we shall consider a negative-binomial distribution. 
The generating function for this
distribution in  full phase space is
\begin{equation}
G^{\mathrm{NBD}}(z)=\left(1+\frac{\bar N}{k}(1-z)\right)^{-k},
\label{15}
\end{equation}
where $\bar N$ represents the average number of particles in 
full phase space and $k$ is a free parameter. Since they describe full
phase space,
both constants of course are $\delta$-independent.  
After the composition (\ref{8}), we obtain the generating
function for the negative-binomial distribution in interval $\delta$ for the
case of statistical phase-space fluctuations 
\begin{equation}
G^{\mathrm{NBD}}(z)=\left(1+\frac{p\bar N}{k}(1-z)\right)^{-k}.
\label{155}
\end{equation} 
Here, $k$ is the same $\delta$-independent constant as in (\ref{15}).
For this distribution, the BPs (\ref{2}) have the following
form
\begin{equation}
\eta_q^{\mathrm{stat}}=\eta_q^{\mathrm{NBD}}=\frac{k+q-1}{k+q-2} \quad ,
\label{14}
\end{equation}
i.e., are $\delta$-independent.

Furthermore, even more complicated distributions exist which lead to
$\delta$-independent BPs for purely statistical fluctuations.
For example, for a convolution of a
number of different negative-binomial multiplicity distributions
\begin{equation}
G^{\mathrm{comp}}(z)=\prod^{\mu}_{s=1}G^{\mathrm{NBD}}_s(z),
\label{16}
\end{equation}
the BPs can be shown not to depend on the interval size $\delta$.

Let us note that dynamical fluctuations  may be
introduced into a model phenomenologically in the form of a projection
(in analogy to (\ref{9})), if we require that for  a subsample of
fixed multiplicity $N$, the phase-space distribution differs from
a positive binomial (so-called bunching projection method \cite{zp20}).
Another way to introduce dynamical fluctuations is by a two-projection
method in which a two-step cluster mechanism with a generating function
for full phase space is postulated in the form of a 
composition of two different 
generating functions. We, therefore, can apply a projection
with two positive-binomial distributions, one for each stage
(for the NBD (\ref{15}) see \cite{zp24}, a general case is 
described in \cite{zp25}).
However, for this method only a monofractal behavior of 
intermittent fluctuations is characteristic. Therefore, 
as shown in \cite{zp20},  for  multifractality it is necessary to use
the bunching projection for both stages, cluster production,
and decay.

\subsection{GHP counting topology}

Now let us illustrate the behavior of the BPs (\ref{3we}) and (\ref{7we})
in  the case of purely  independent phase-space  distribution, 
using the GHP counting topology.
As we have noted in Sect.~3, if the full-phase-space multiplicity
distribution is not Poissonian, then the values of the 
generalized  BPs are not
equal to unity.
 
An event sample is obtained with a random event generator
\footnote{To generate $N$ 
independent points for each event, 
we use the generator NRAN for  uniformly distributed pseudo-random
numbers (CERN Program Library).} in the following way: 
For a given event of multiplicity $N$ in 
full phase space, we generate $N$ independent  
pseudo-random points in the ``phase space'' $0<x<1$. 
After that, we simulate the 
distribution for multiplicity $N$.

In Figs.~2 and 3 we present the $M=1/\epsilon$ -behavior of
differential BPs for two-particle spikes $\chi_q^{\mathrm{stat}}(1/M,2)$ and
integral BPs $\chi_q^{\mathrm{stat}}(1/M)$ for purely independent
production of particles in the phase space $x$.
The total number of events is $10^6$.
Since the behavior of statistical fluctuations
depends on the full-phase-space multiplicity distribution,
we have considered the generalized BPs for the following cases:

1) $N$ is fixed for all events ($N=21$). This case is
shown by open squares in the figures.
Here, $\chi_{q}^{\mathrm{stat}}(1/M ,2)<1$ 
and $\chi_{q}^{\mathrm{stat}}(1/M)<1$.
Such an anti-bunching effect
is a consequence of trivial negative correlations that are present, when 
the probability of  finding a spike is less if
another spike has already been found.

2) $N$ is distributed according to a Poissonian distribution
with average $\bar{N}=21$ (closed squares).
As expected, the values of the
bunching parameters are equal to unity.

3) In order to study a more realistic case, 
we generated the distribution for 
charged-hadron multiplicity $N$ in
full phase space according to JETSET 7.4 PS. 
To investigate the sensitivity of the BPs to various forms of 
single-particle distribution, 
we consider two different cases. In the first case,
the phase-space density is uniform, i.e. $\rho(x)=dn/dx=const$
(open circles in the figures). For the second case, the
phase-space density has the strongly non-uniform shape
$\rho(x)=const\,(1+x)^{-6}$
(closed circles)\footnote{Such a single-particle inclusive density can 
easily be obtained as the product
of two generators for uniformly distributed
pseudo-random numbers.}. 
As we see from Figs.~2 and 3, the generalized distance-measure BPs 
have values larger than unity. 
Hence, the corresponding spike multiplicity
distributions are broader than a Poissonian distribution. 

The most important feature of the 
generalized distance-measure BPs considered here is that,
in the case of  independent
production of particles, they are approximately independent of the 
spike size $\epsilon$. Only for the full-phase-space multiplicity
distribution generated by JETSET 7.4 PS is, a small rise of
the generalized BPs visible for not very large $M$.
In contrast to the bin-splitting definitions of BPs, 
the generalized BPs probably rise with decreasing $\epsilon$
even for  very small values of $\epsilon$  due to
the deviation in full-phase-space multiplicity distribution from
a Poissonian distribution. However, to derive
an exact conclusion on the full-phase-space dependence
of generalized BPs, more investigation is
needed, since statistical
errors in the figures are comparable with the size of the symbols.

Figs.~2 and 3 show that the result obtained for JETSET 7.4 seems to
be independent of the form of the single-particle density.
It  is important to note that 
a non-uniform phase-space density (closed circles) leads
to a more stable result for the $M$-dependence and   
significantly reduces the statistical error. 

Fig.~4 shows the behavior of $\eta_q(1/M)$ (\ref{2zpo}) for
$q=3,4$ as a function of $\epsilon=1/M$ for the case of
a Poissonian full-phase-space multiplicity distribution with
average $\bar{N}=21$. 
The total number of events is the same as that for Figs.~2 and 3.
The independent particle distribution over phase space is
simulated as for  Figs.~2 and 3. 
Fig.~4 demonstrates
that the corresponding multiplicity
distribution $\tilde P_n(\epsilon)$ is  narrower than
Poisson ($\eta_q(1/M)<1$), even if the particles are 
produced independently of each other. However, the main deficiency
of definition (\ref{2zpo}) lies in the insufficient use of
statistics available. This leads to large statistical errors
for large $M$. The calculation of $q=5$ and $q=6$ for $M>100-200$, therefore,
was found
impossible due to limited statistics (not shown).

The subject of the behavior of generalized BPs  is
complex and, probably, must be solved separately
for each particular type of BPs with
a given definition of spike size, for a given 
multiplicity distribution of particles in full phase space.
However, any $\epsilon$-dependence of the BPs
for  purely statistical fluctuation 
due to full-phase-space fluctuations can 
be completely suppressed by using
$1/\chi_q^{\mathrm{stat}}$ or $1/\eta_q^{\mathrm{stat}}$ 
as a correction factor.
After the correction procedure,
any deviation in the  behavior of the corrected generalized BPs from unity
can be interpreted
as being due to the presence of genuine local multiplicity fluctuations.

\section{Local fluctuations in the JETSET 7.4 model}

A widely used means to study general features of hadronic final-state
fluctuations is to simulate hadronic events according to Monte-Carlo models.
Below we will consider the behavior of BPs for hadrons produced in
$\mathrm{e}^+\mathrm{e}^-$-annihilation at $91.2$ GeV 
using the JETSET 7.4 PS model.

To study local fluctuations in this model, we use the bin-averaged
BPs (\ref{3}) with horizontal normalization. The azimuthal angle $\varphi$,
calculated with respect to the beam axis, is used as a phase-space
variable. Since there is no preferred direction for
hadrons, the event averaged distribution in $\varphi$ is uniform.

Fig.~5a shows for four different ranks $q$ the value of $\eta_q$ 
as a function of $M$, where $M=2\pi /\delta\varphi$ is the number of
partitions of the full azimuthal angle $2\pi$.
The number of
events generated is 750k. From this figure it follows that there
is a power-like behavior of the second-order BP, but
all higher-order BPs tend to decrease with increasing $M$. 
Such an anti-bunching
trend for higher-order BPs is the result of jet
formation combined with energy-momentum conservation:
particles belonging to different jets are separated by phase space.

In Fig.~5b we present the $M$-dependence of the BPs in azimuthal angle,
but now calculated with respect to the thrust axis.
Since the distribution for this kind of measurement is far from flat,
the transformation \cite{zp12,zp13} of azimuthal-angle
variable to a new cumulative variable with flat
single-particle density was performed before the calculation of BPs.
Fig.~5b shows a
power-law trend in the behavior of all BPs studied, without any
visible saturation for large $M$, as is usually seen for NFMs in
one-dimensional variables. We can conclude
that the multifractal structure of intermittency is
an inherent feature of fluctuations in the azimuthal angle defined
with respect to the thrust axis. This  means  that multifractality is
mainly a feature of fluctuations inside jets, rather than a property
of fluctuations in the $\varphi$ variable 
defined with respect to the beam.

Note that for small $M$, the behavior of the BPs is not  meaningful:
as we have seen  in the previous  section, in the domain
$M\le 10 - 20$ the value of the BPs can
be affected by statistical fluctuations.
In this case, an $M$-dependence of BPs can occur
even without any dynamical reason.
In addition, for small $M$, as is the case for  NFMs, BPs are affected by
the large-scale structure of fluctuations for which energy-momentum
constraints are characteristic.

To compare the result obtained  with NFMs, 
we present in Fig.~6a,b the
behavior of NFMs as a function $M$, where we use the azimuthal angle 
$\varphi$ calculated with respect to the beam axis (Fig.~6a) and the
thrust axis  (Fig.~6b). Both calculations show qualitatively 
the same trend and it is very difficult to derive a conclusion on
a different behavior  of these two intermittent samples.

The same conclusion has been derived in \cite{zp26}, 
where a theoretical local-fluctuation 
model was studied  with the help of both NFMs and BPs.
It has been shown that two very different model samples 
can lead to rather similar power-law behavior of NFMs, while the 
BPs show a different trend. This means, in fact,
that the NFMs are not sensitive to the details in the structure
of  intermittent fluctuations. 
The  good agreement between experimental behavior of NFMs
and Monte-Carlo predictions, as claimed recently \cite{zp27,zp28,zp29}, 
therefore, cannot provide
a final proof of the similarity between experimental intermittent samples
and samples generated by Monte-Carlo models in ever smaller phase-space
intervals.

To demonstrate the behavior of generalized BPs,
we use the squared four-momentum difference between two charged particles
$Q^2_{12}=-(p_1 -p_2)^2$ as a distance measure. 
Fig.~7 shows the behavior of integral
$\chi_q(Q^2_{12})$ (closed circles) and differential $\chi_q(Q^2_{12}, 2)$
(open circles) bunching parameters. 
The dashed line represents the behavior of these BPs
in the case of a Poissonian distribution.
Both kinds of BPs rise with decreasing $Q^2_{12}$.
This corresponds to a strong bunching effect. The saturation and 
downward bending of
the second-order BPs at small $Q^2_{12}$ is 
caused by the influence of
resonances at intermediate $Q^2_{12}$. We have verified that 
such behavior  
disappears for like-charged particle combinations (not shown).
The latter observation is very important, since the rise
of BPs for identical pions with decreasing $Q^2_{12}$ 
can be attributed to Bose-Einstein
correlations.

It is quite remarkable that the value of $\chi_q(Q^2_{12})$  is
always larger than $\chi_q(Q^2_{12}, 2)$, especially for not very
small $Q^2_{12}$. For small $Q^2_{12}$, both definitions
of BPs show the same trend and have similar values. 
The reason for such a 
similarity becomes clear when one realizes that the 
integral
BPs include the contribution from 
two-particle spikes. For small interparticle distances,
the integral BPs are then dominated by two-particle spikes. 

For large $Q^2_{12}$, 
the contribution of
many-particle spikes to $\chi_q(Q^2_{12})$ is more sizable. 
In such a case, the integral
BPs are more sensitive,  than are the differential ones,
to jet events. This is due to the fact
that jets can contribute to $\chi_q(Q^2_{12}, 2)$
only if they contain exactly two charged particles in each jet.
In contrast, the integral BPs are effected by jets
with a different number of particles. For example, for large $Q^2_{12}$, 
the second-order integral BP is strongly influenced by two-jet events, 
the third-order BP is sensitive to both two- and three-jet events
and so on.

\section{Conclusions}

Intermittency, as originally considered for particle physics 
by Bia\l as and Peschanski \cite{zp1},
is a term borrowed from turbulence theory, 
as are most of the mathematical techniques used in this field,
which is why  intermittency was formulated in  terms 
of continuous particle densities. In that approach,
a convolution was assumed of an underlying dynamical density distribution with
multi-Poissonian statistical noise. For such a situation, 
the method of removing  statistical noise by the normalized factorial
moments  follows immediately.

However,
the problem of intermittent dynamical fluctuations may, in principle,
also be  described in  terms of 
bunching parameters.
As is the case for bin-averaged
normalized factorial moments, the bin-averaged BPs  remove 
the influence of Poissonian statistical noise for small $\delta$ and 
become $\delta$-independent constants 
if fluctuations have only statistical
origin. Furthermore, definitions of the BPs are given which 
can be used for the study of fluctuations in various
phase-space variables, without any artificial binning of phase space.
This property is very important for the investigation of  Bose-Einstein
correlations and  resonance decays. 

As  mentioned in the introduction, one of the most important
properties of the BPs is that these quantities are not affected by
the experimental statistical bias which arises in NFMs when the bin size 
becomes very small.
Of course, the limitation in number of experimental events  leads  to
an increase of the statistical errors with decreasing $\delta$ (or $\epsilon$)
for lower-order BPs and
to the failure to calculate  higher-order BPs.
In  contrast, the NFMs tend to be depressed 
at very small $\delta$ as compared to their 
values expected for an infinite sample \cite{zp7}.

Moreover, in studying intermittent fluctuations, 
there is a trivial tendency in the
behavior of the NFMs: the higher the order of the NFM,
the larger is its value for a given  $\delta $ (or $\epsilon$).
On the contrary, the high-order BPs, in principle, can have
any dependence on $\delta$ (or $\epsilon$), i.e., the possible 
behavior of the BPs has a larger number of ``degrees of freedom''.
This observation provides tools for a better understanding of the  
differences between samples with approximately 
the same  power-law behavior of the NFMs
and a selective study of fluctuations in terms of  
different types of spikes.   

The last point has a primary importance for the investigation of
local multiparticle fluctuations inside jets.
The behavior of
NFMs is qualitatively the same \cite{zp27,zp28,zp29} 
for variables defined with respect to the beam axis 
and with respect to the sphericity axis.
The information content of these measurements, however, is rather different.
The spikes dominating the distributions in variables defined with 
respect to the beam axis are due to the jets produced in a given event.
Such spikes are separated in  phase space
because of energy-momentum conservation.
This trivial effect always dramatically affects the observed behavior 
of local quantities measured
in variables with respect to the beam axis.  
On the other hand, any local measurements of phase-space distributions
in variables defined with respect to the sphericity or thrust axes 
mainly reflect
the physical content of fluctuations that arise due to underlying
stages (perturbative and fragmentation stages, resonance decays,
Bose-Einstein interference) of multihadron production inside jets.
Since the behavior of NFMs is not sensitive to the definition of a
preferred axis, it is quite difficult to determine the physical nature of
the intermittent signal observed for the two cases mentioned.

As we have seen, the different definitions of 
generalized BPs merely reflect the freedom of choice of
event configurations. From the experimental point
of view, this is very handy, since we can choose a 
form of  BPs optimized according to a given statistics of
an experiment and according to the aims of the investigation.

We hope that the use of BPs 
will be  useful 
for the investigation of  details in
the multifractal behavior of particle spectra,
where it is important to find
and to study the contributions from different multiparticle
clusters  and to compare  theoretical or model
multiplicity distributions with the experimental data.

\newpage
\bigskip
Acknowledgments
\medskip

This work is part of the research program of the ``Stichting voor
Fundamenteel Onderzoek der Materie (FOM)'', which
is financially supported by the ``Nederlandse Organisatie voor  
Wetenschappelijk Onderzoek (NWO)''. 
The research  of S.V.C. was supported in part by
the International Soros Science Educational Program.

\newpage
\begin{figure}
\begin{center}\mbox{\epsfig{file=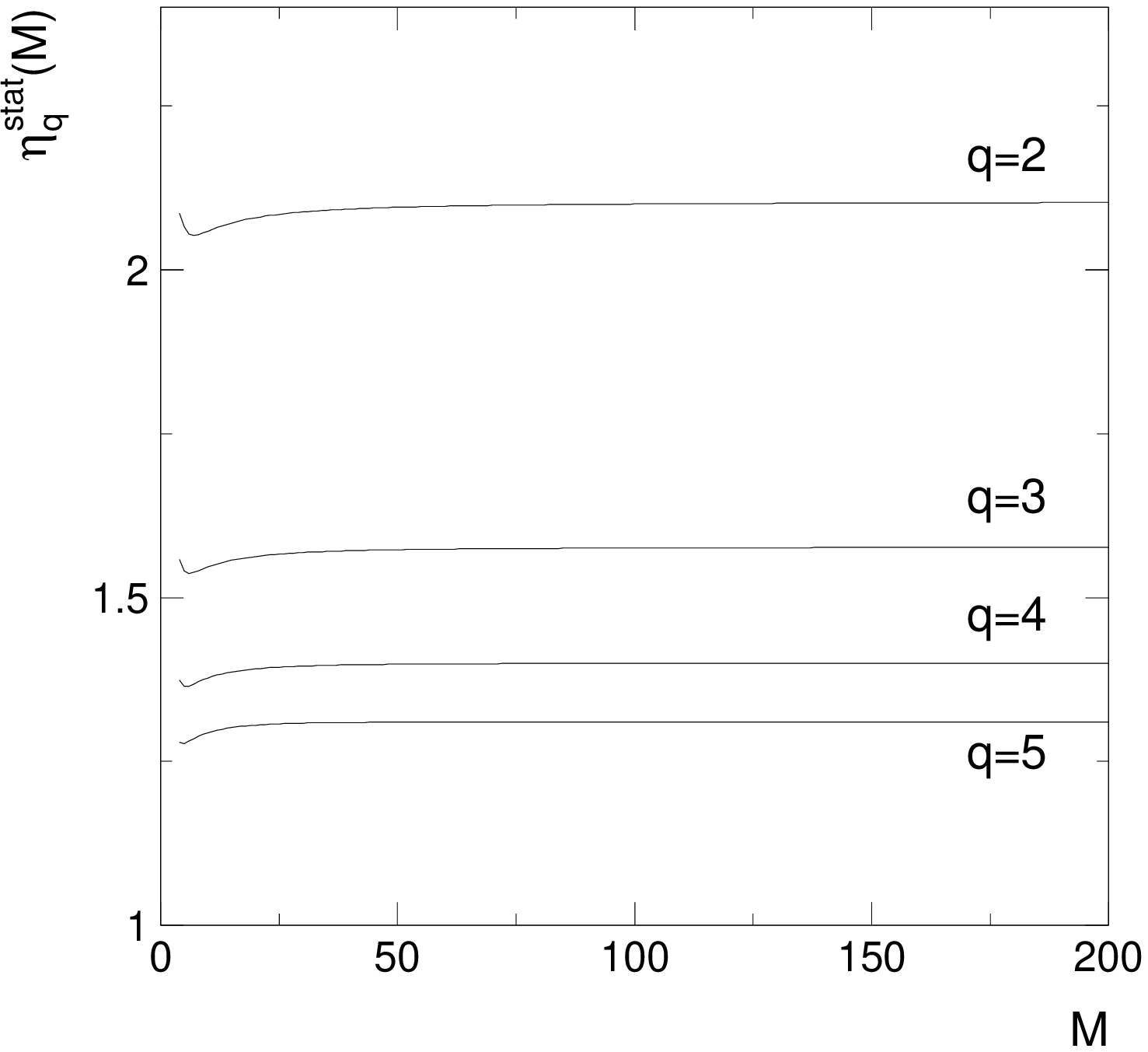,height=12cm}}
\end{center}
\vspace{1cm}
{\bf Fig.~1:} The BPs as a function of $M$
in the case of statistical phase-space fluctuations. Here we use an analytical
description of the phase-space distribution 
in the form of a positive-binomial distribution 
and simulate the multiplicity
distribution for full phase space by JETSET 7.4.
\end{figure}

\newpage
\begin{figure}
\begin{center}\mbox{\epsfig{file=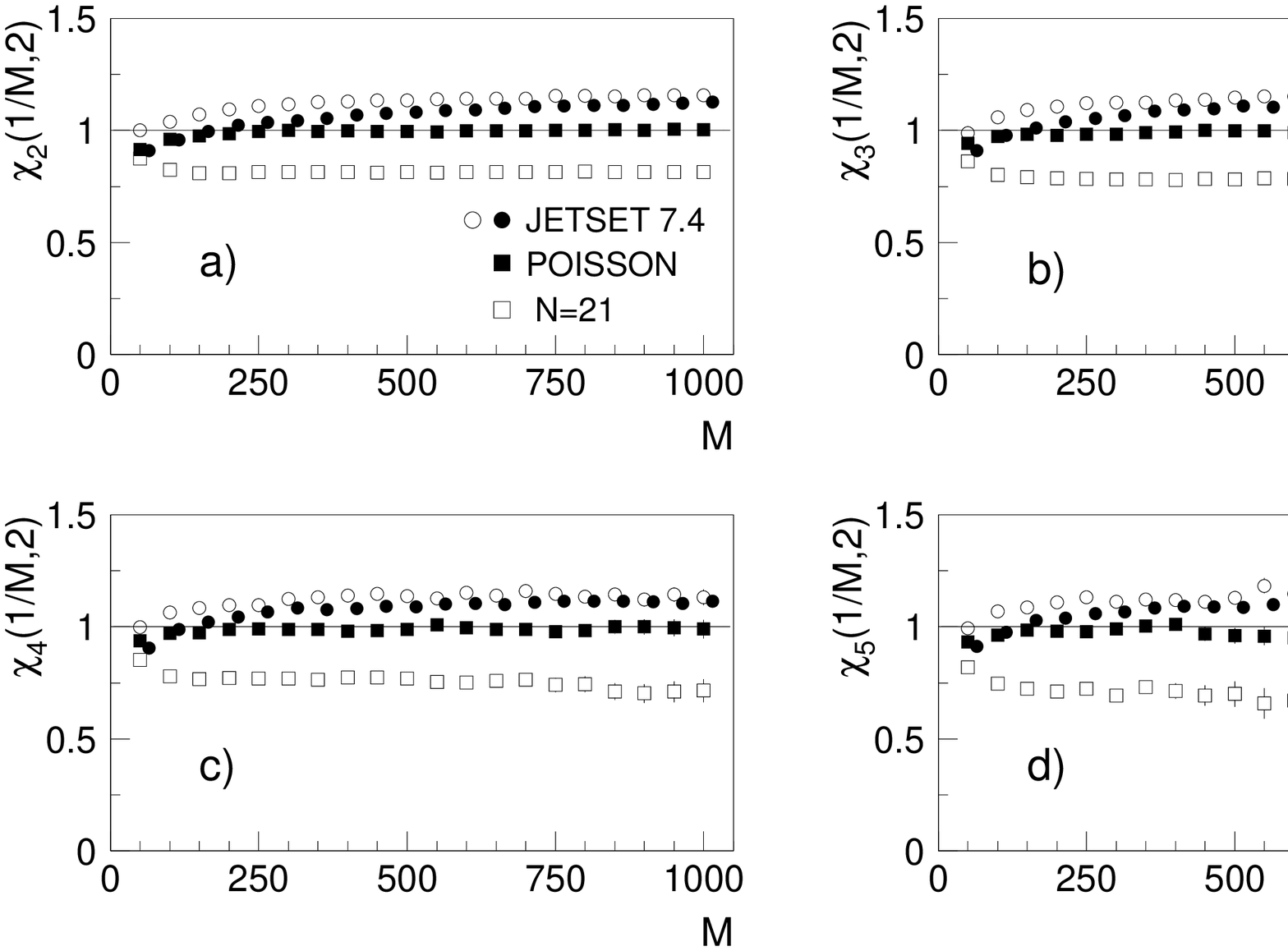,height=8.6cm, width=14.5cm}}
\end{center}

\vspace{0.1cm}
{\bf Fig.~2:}
The values of differential BPs $\chi_{q}^{\mathrm{stat}}(1/M, 2)$
as a function of $M=1/\epsilon$
in the case of statistical  fluctuations.
The open circles represent the
uniform single-particle density $\rho(x)=const$ and the closed 
circles correspond to non-uniform density $\rho(x)=const\,(1+x)^{-6}$.

\vspace{0.1cm}

\begin{center}\mbox{\epsfig{file=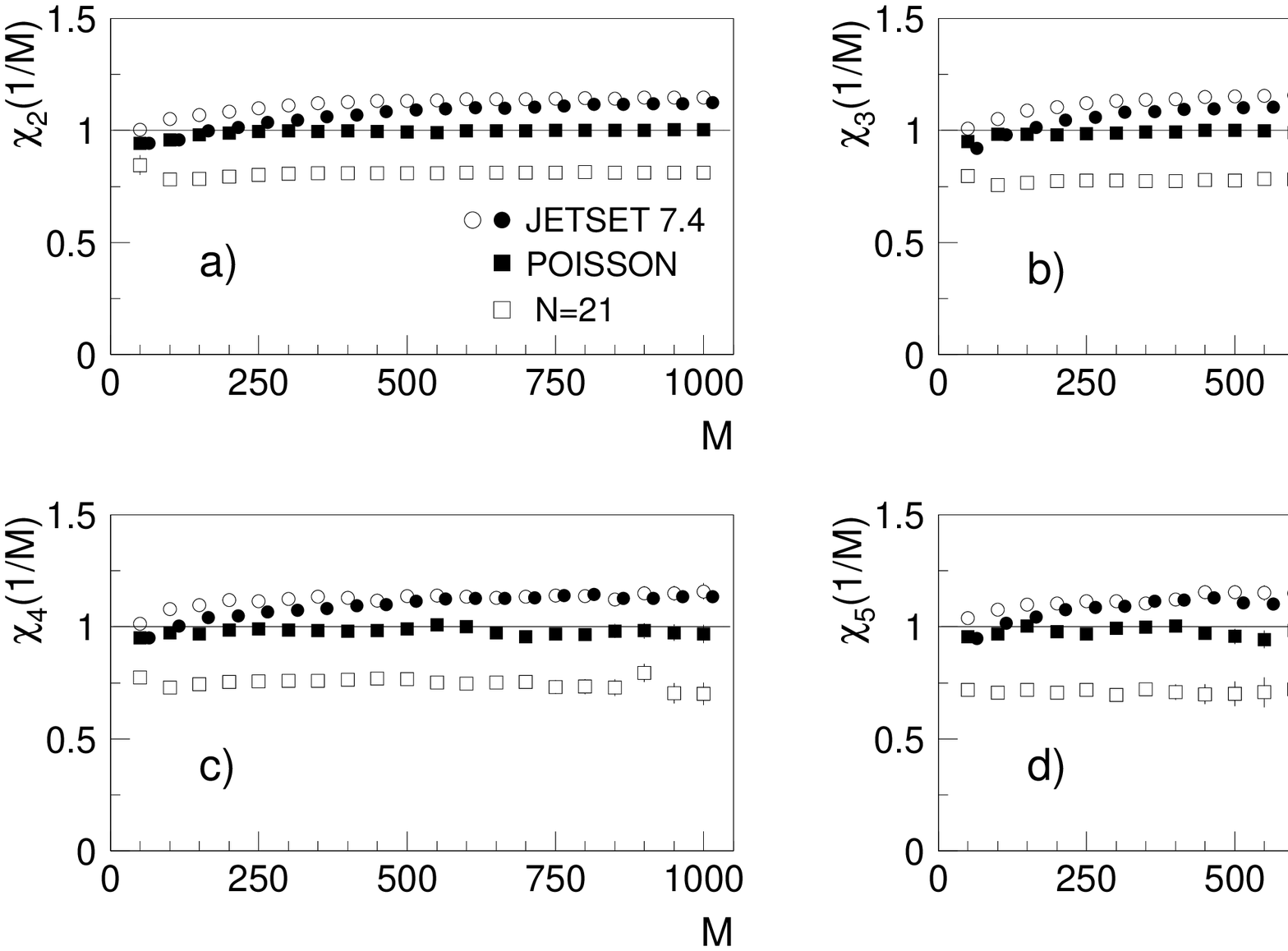,height=8.6cm, width=14.5cm}}
\end{center}

\vspace{0.1cm}
{\bf Fig.~3:}
The values of integral BPs $\chi_{q}^{\mathrm{stat}}(1/M)$
as a function of $M=1/\epsilon $
in the case of statistical  fluctuations.
\end{figure}

\newpage
\begin{figure}
\begin{center}\mbox{\epsfig{file=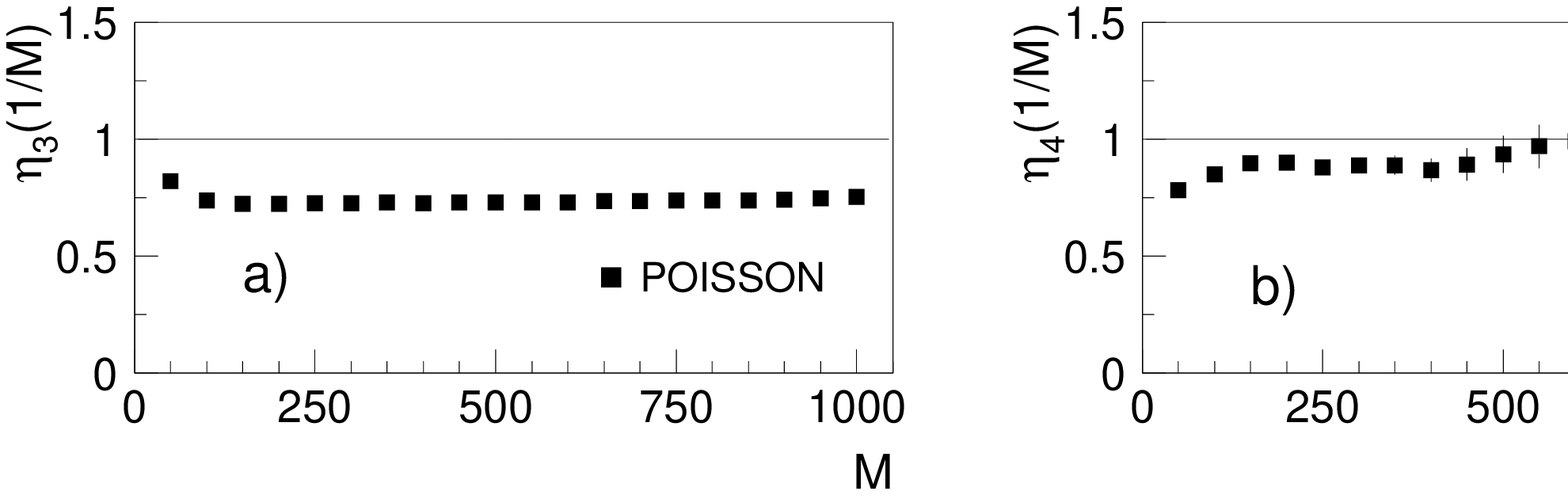, height=5.0cm}}
\end{center}
\vspace{0.1cm}
{\bf Fig.~4:}
The values of  BPs $\eta_{q}^{\mathrm{stat}}(1/M)$ (\ref{2zpo})
as a function of $M=1/\epsilon$
in the case of statistical  fluctuations.
\end{figure}

\newpage
\begin{figure}
\begin{center}\mbox{\epsfig{file=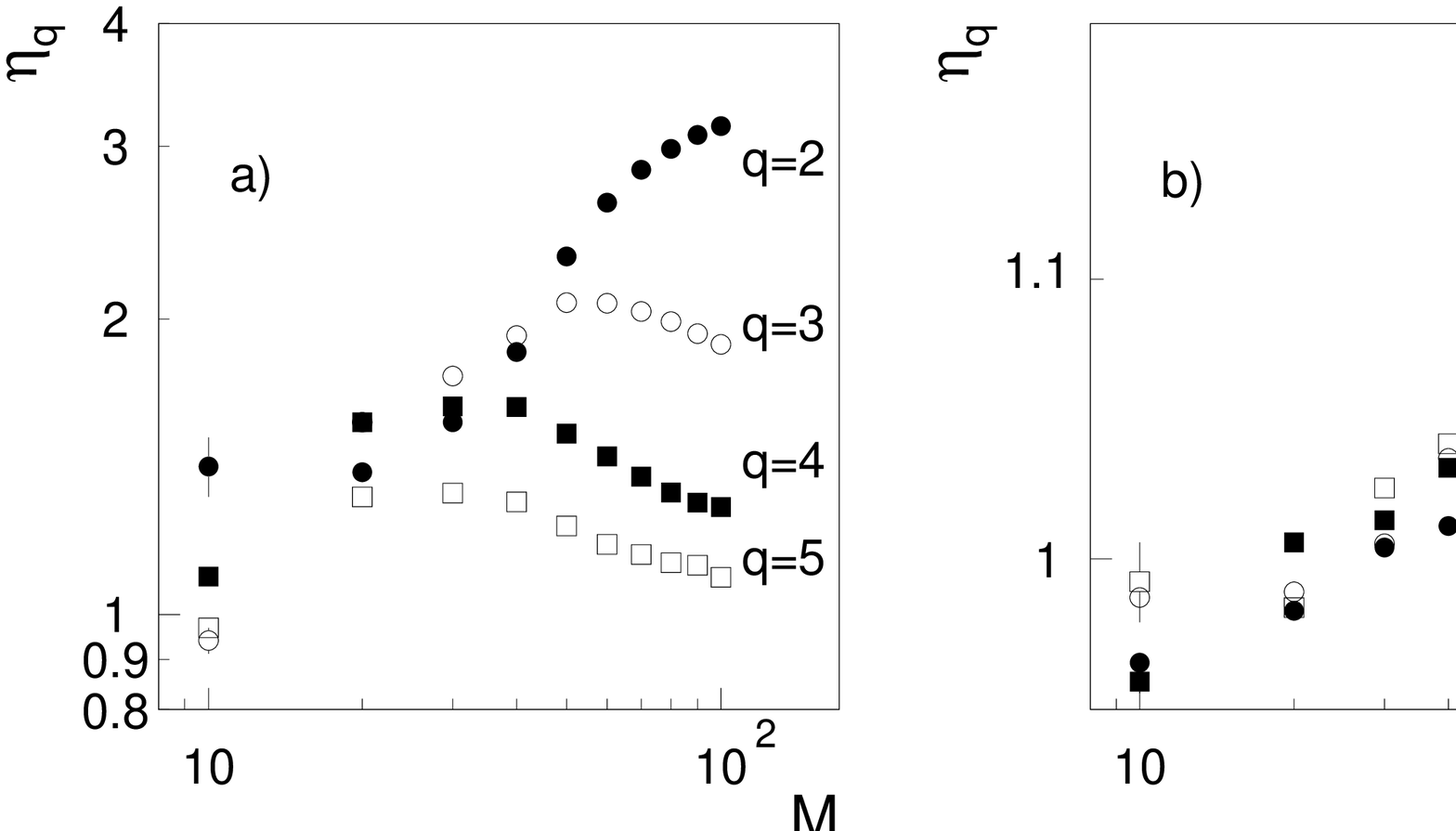,height=6.5cm, width=15.3cm}}
\end{center}

\vspace{0.4cm}
{\bf Fig.~5:} 
BPs as a function of the number of bins in the  azimuthal angle $\varphi$
defined with respect to the {\bf a)} beam axis  and {\bf b)} thrust axis.
(JETSET 7.4 PS)

\vspace{1.0cm}

\begin{center}\mbox{\epsfig{file=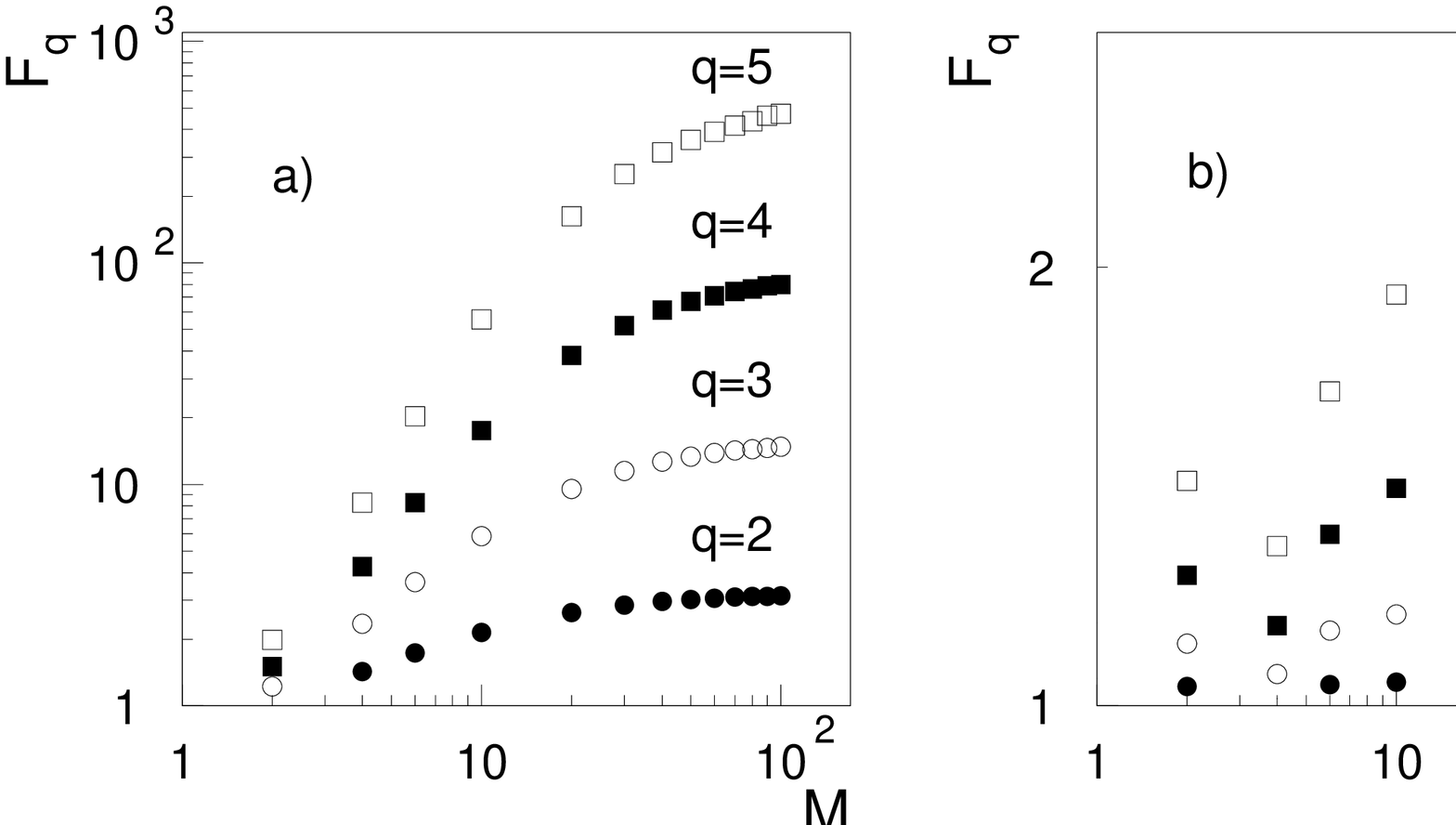,height=6.5cm, width=15.3cm}}
\end{center}

\vspace{0.4cm}
{\bf Fig.~6:} 
NFMs as a function of the number of bins in the azimuthal angle $\varphi$
defined with respect to the {\bf a)}
beam axis and {\bf b)}thrust  axis (JETSET 7.4 PS model).
\end{figure}

\newpage
\begin{figure}
\begin{center}\mbox{\epsfig{file=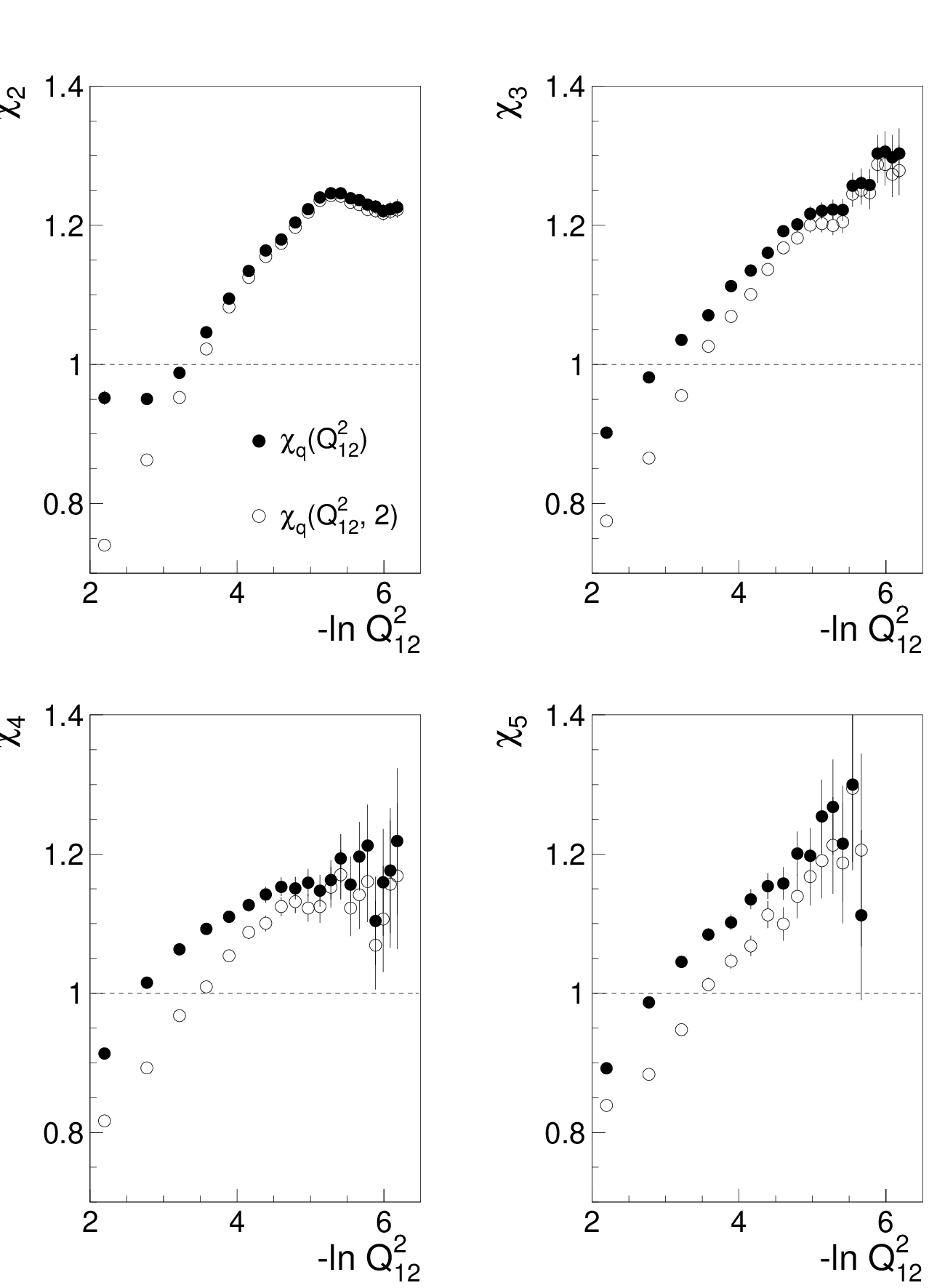,height=15.0cm }}
\end{center}

\vspace{0.4cm}
{\bf Fig.~7:}
Integral (closed symbols) and differential (open symbols)
BPs as a function of the squared four-momentum difference $Q^2_{12}$ 
between two charged particles, calculated in the  JETSET 7.4 PS model. 
\end{figure}

\end{document}